\pdfoutput=1
\documentclass[12pt]{article}
\usepackage[UKenglish]{babel}
\usepackage[UKenglish]{isodate}
\makeatletter
\renewcommand*{\@fnsymbol}[1]{\ensuremath{\ifcase#1\or \dagger \or * \or ** \else\@ctrerr\fi}}
\makeatother
\usepackage{amsmath}
\usepackage{bm}
\usepackage{graphicx}
\usepackage{hyperref}
\title{\fontsize{19}{22.8}\selectfont Fresnel versus Kummer surfaces: geometrical optics in dispersionless linear (meta)materials and vacuum\thanks{\textit{Electromagnetic Spacetimes}, Wolfgang Pauli Institute, Vienna, 19--23 November 2012.}}	
\author{\underline{Alberto Favaro}\thanks{E-mail: favaro@thp.uni-koeln.de},\\
Inst.\ Theor.\ Phys., Univ.\ of Cologne, Germany\\
\ \\
Friedrich W.\ Hehl\thanks{E-mail: hehl@thp.uni-koeln.de},\\
Inst.\ Theor.\ Phys., Univ.\ of Cologne, Germany \textit{and}\\
Dept.\ Phys.\ \& Astron., Univ.\ of Missouri, Columbia, USA
}
\cleanlookdateon
\date{\vspace{5pt}2 November 2012}
\begin{document}
\maketitle

\textit{Geometrical optics} describes, with good accuracy, the propagation of high-frequency plane waves through an electromagnetic medium. Under such approximation, the behaviour of the electromagnetic fields is characterised by just three quantities: the temporal frequency $\omega$, the spatial wave (co)vector $k$, and the polarisation (co)vector $a$. Numerous key properties of a given optical medium are determined by the \textit{Fresnel} surface, which is the visual counterpart of the equation relating $\omega$ and $k$. For instance, the propagation of electromagnetic waves in a uniaxial crystal, such as calcite, is represented by two light-cones. \textit{Kummer}, whilst analysing quadratic line complexes as models for light rays in an optical apparatus, discovered in the framework of projective geometry a \textit{quartic} surface that is linked to the Fresnel one. Given an arbitrary dispersionless linear (meta)material or vacuum, we aim to establish whether the resulting Fresnel surface is equivalent to, or is more general than, a Kummer \nolinebreak surface. 

\begin{figure}
  \centering
    \includegraphics[width=0.7\textwidth]{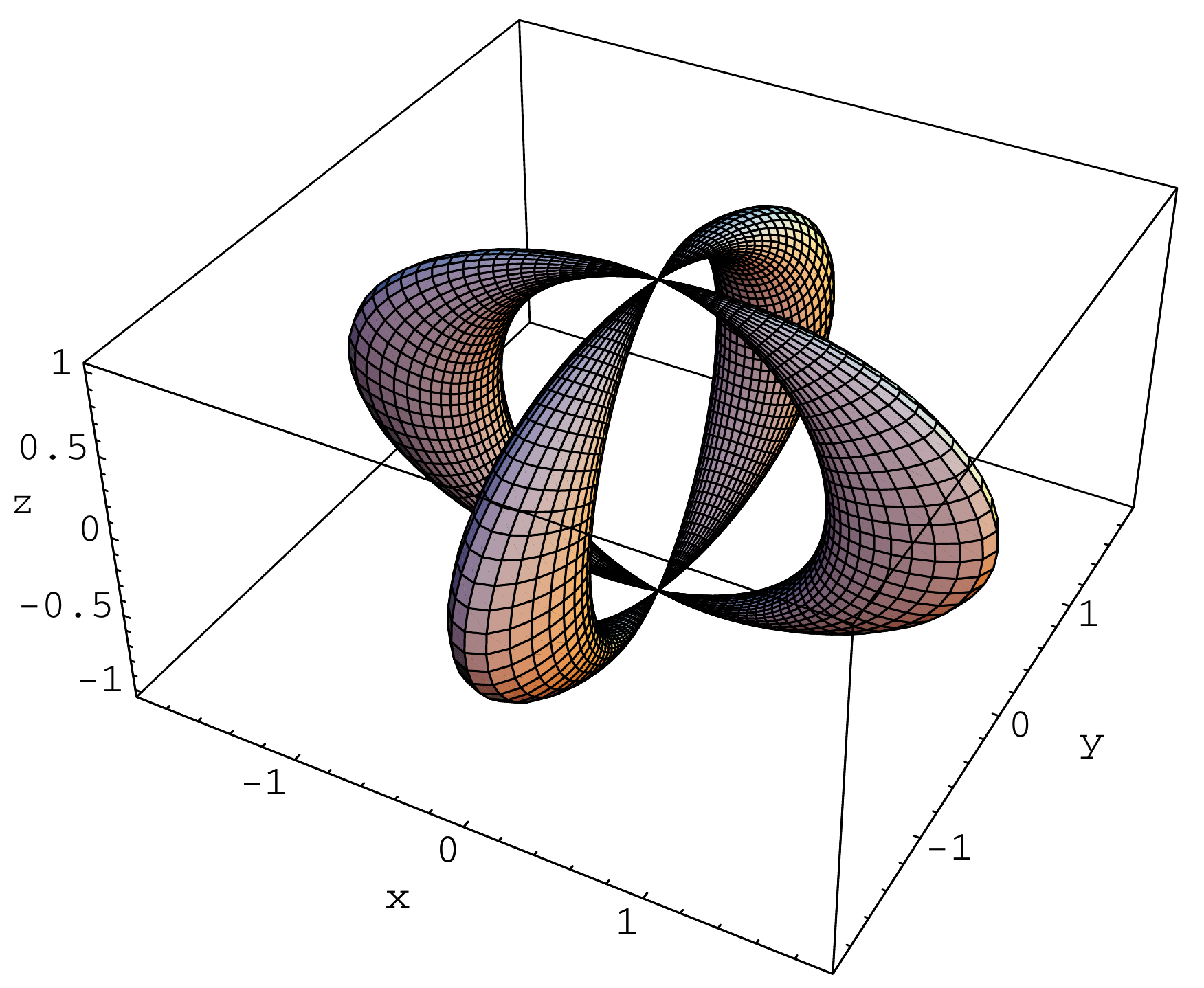}
\caption{\label{fig:skewon}Does this skewonic medium really have a Fresnel=Kummer surface? Taken from: Obukhov and Hehl (2004).}
\end{figure}

In the 1905 book \cite{hudson1905}, Hudson examines the relation between $\omega$ and $k$ originating from a wide family of dielectric crystals ($\varepsilon\!=$anisotropic, but ${\mu\!=\!\mu}_0$), and deduces that the Fresnel surface is a restricted case of the Kummer one. Four years later, this conclusion is rectified by Bateman \cite{bateman1910}. Starting from a more general ansatz, to wit, from a dispersionless linear medium that is constrained by one symmetry requirement alone, he appears to demonstrate that the Fresnel surface coincides with the Kummer one. A number of recent works investigate the properties of those (meta)materials and vacuum geometries for which the additional symmetry condition does not hold true \cite{lindell2012, itin2009, dahl2012, favaro2012}. One is thus motivated to verify whether such optical media, endowed a non-zero \textit{skewon} component, still give rise to a Fresnel surface that is equivalent to a Kummer one. It may of course happen that the propagation of light in some dispersionless and linear (meta)materials or theories of vacuum is not described by a Kummer surface, cf.\ Figure \ref{fig:skewon}. One would then proceed and examine K3 surfaces and Calabi-Yau manifolds of increasing \nolinebreak generality. 

The article \cite{bateman1910} by Bateman introduces several notions and techniques which became of common use only in recent years. Moreover, it describes the theory of geometrical optics from a novel and potentially fruitful angle. Bateman investigates the propagation of light in a medium such that, not only the permittivity and permeability are anisotropic, but also the magneto-electric response is non-zero. As mentioned earlier, besides the usual requirements of linearity and zero dispersion, the constitutive law discussed in \cite{bateman1910} satisfies
\begin{equation}
  \chi^{\alpha\beta\mu\nu}=\chi^{\mu\nu\alpha\beta} \qquad \mbox{(no skewon)}.
\end{equation}
Here, $\chi^{\alpha\beta\mu\nu}$ is the structure tensor that links the field excitation ${\check{H}^{\alpha\beta}\!\sim\!(\mathcal{D},\mathcal{H})}$ and the field strength ${F_{\alpha\beta}\!\sim\!(E,B)}$ according to ${\check{H}^{\alpha\beta}\!=\!\frac{1}{2}\chi^{\alpha\beta\mu\nu}F_{\mu\nu}}$. It is important to observe that the medium is subject to no further restrictions. More in detail, the component of the electromagnetic response that is singled out by the index alternation  $\chi^{[\alpha\beta\mu\nu]}$ need not vanish. As a result of this choice, Bateman is seemingly the first to introduce an electromagnetic medium that has a non-zero \textit{axion} part. He is also one of the few authors that relate geometrical optics and the tracing of lines in a 3-dimensional projective space.

\bibliography{vienna}
\bibliographystyle{unsrt}
\end{document}